\documentstyle[epsfig]{aipproc}

\begin{document}
\title{The Evolution of Dark-Matter Dominated Cosmological Halos}

\author{Marcelo Alvarez$^*$, Paul R. Shapiro$^*$ and Hugo Martel$^*$}
\address{$^*$Department of Astronomy, University of Texas,
         Austin, TX 78712}

\maketitle

\begin{abstract} Adaptive SPH and N-body simulations were carried out to
study the evolution of the equilibrium structure of dark matter
halos that result from the gravitational instability and fragmentation of
cosmological pancakes. Such halos resemble those formed by
hierarchical clustering from realistic initial conditions in a CDM
universe and, therefore, serve as a test-bed model for
studying halo dynamics.  The dark matter density profile is close to the
universal halo profile identified previously from N-body simulations of
structure formation in CDM, with a total mass and concentration parameter
which grow linearly with scale factor $a$.  When gas is included, this
concentration parameter is slightly larger than the pure N-body result.  
We also find that the dark matter velocity distribution is less isotropic
and more radial than found by N-body simulations of CDM.  
\end{abstract}

{\bf CDM Simulations vs.\ Observed Halos} N-body simulations of
structure
formation from Gaussian-random-noise density fluctuations in a cold dark
matter (CDM) universe have revealed that dark matter
halos possess a universal density profile that diverges as $r^{-\gamma}$
near the center, with $1\leq\gamma\leq2$ \cite{Moore00}\cite{NFW97}.
There exists a discrepancy between these singular density profiles found
in N-body simulations and current observations of the rotation curves of
nearby dwarf galaxies \cite{Moore00} and of strong lensing of background
galaxies by the galaxy cluster CL0024+1654 \cite{Tyson98}\cite{Shapiro00},
which suggest that dark-matter dominated halos of all scales have flat
density
cores, instead.

\begin{figure}[t!] 
\centerline{\epsfig{file=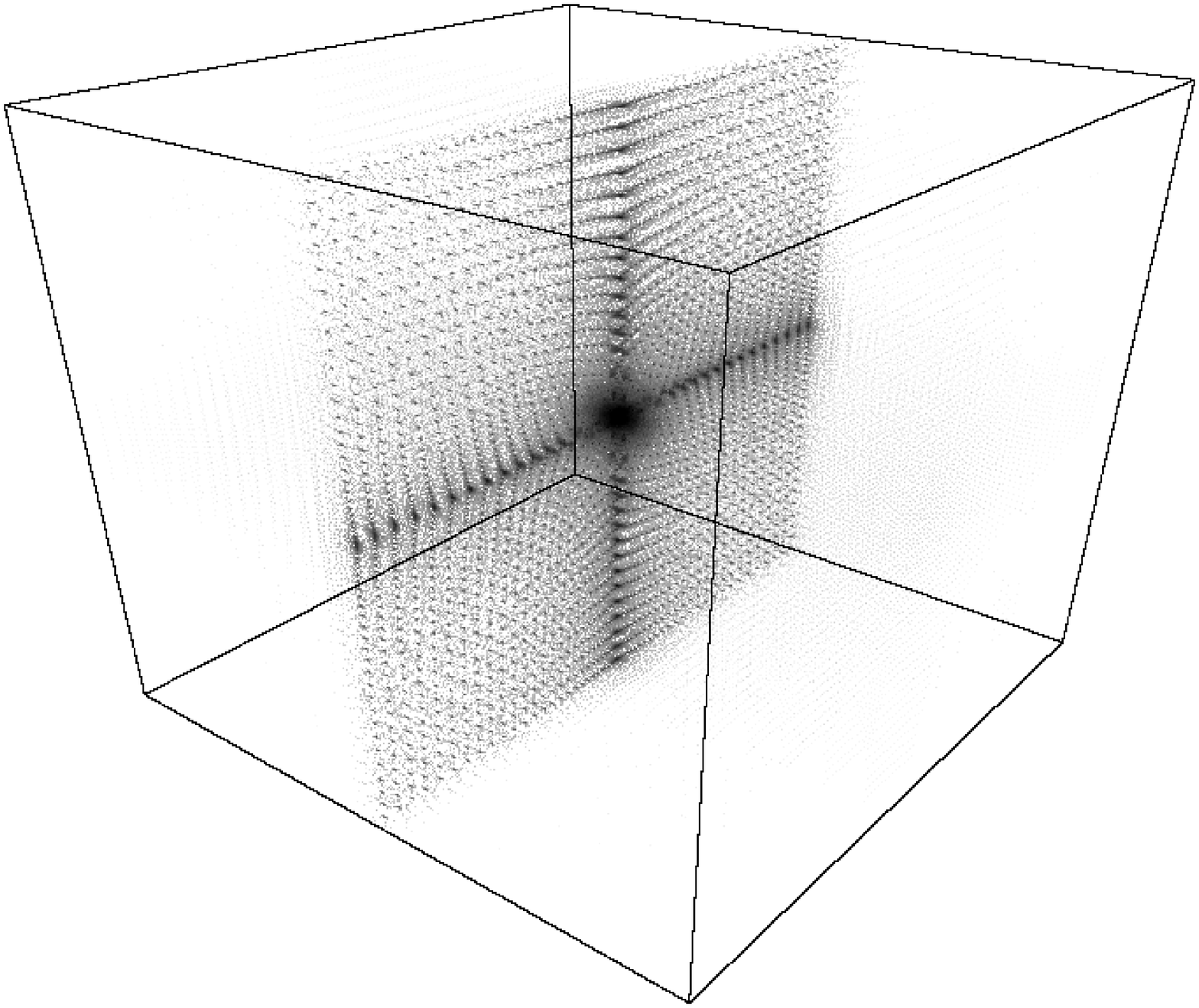,width=2.6in}
\epsfig{file=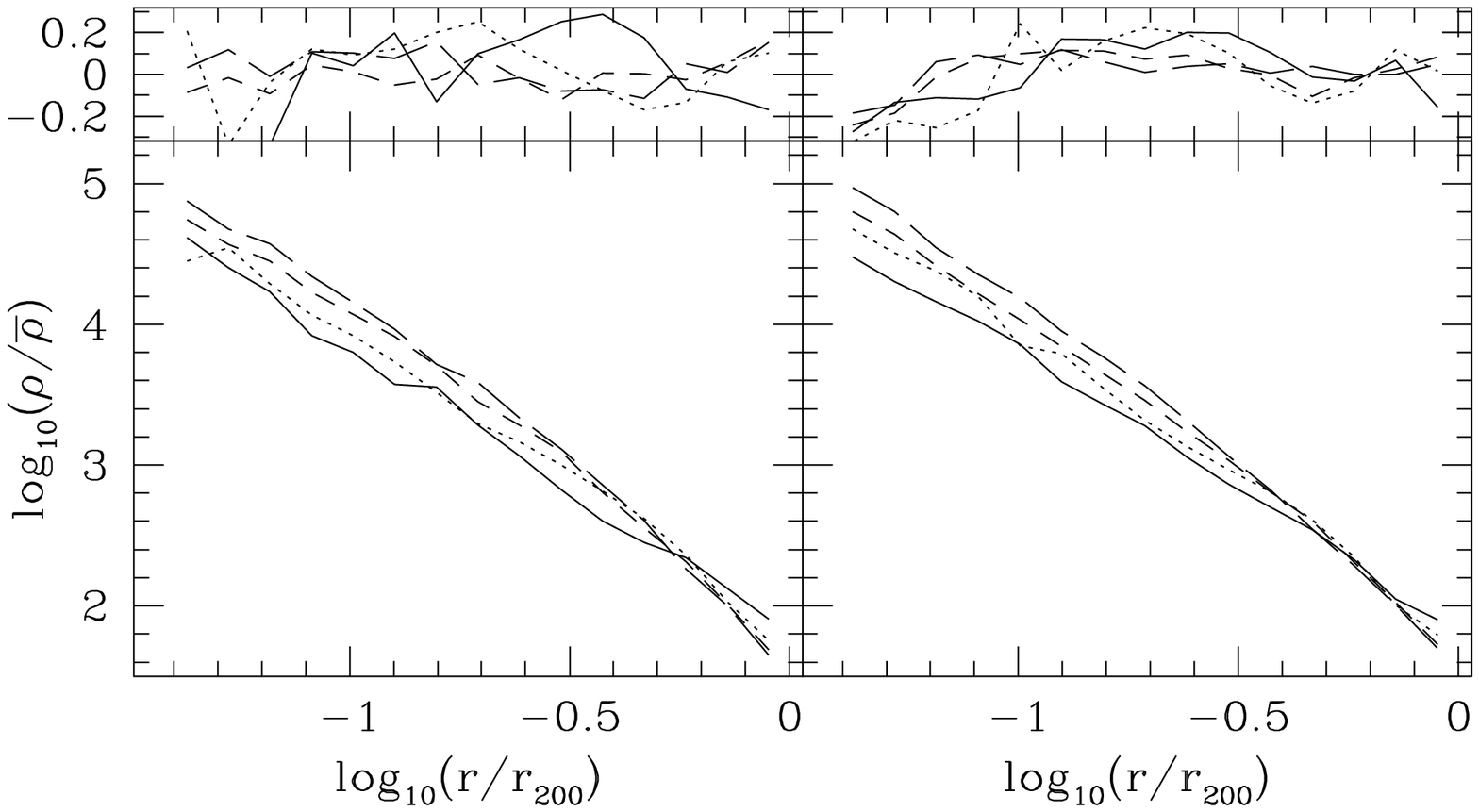,width=4.0in}}
\vspace{10pt}

\caption{(left) Dark matter density field at $a/a_c=3$.  (middle) Density
profile of the dark matter halo as simulated without gas at four different
scale factors, $a/a_c=3$ (solid), 4 (dotted), 5 (short dash), and 7 (long
dash).  Shown above are fractional deviations
$(\rho_{NFW}-\rho)/\rho_{NFW}$ from best-fit NFW
profile for each epoch. (right) Same as middle, but for DM halo simulated
with gas+DM. }

\label{fig1}
\end{figure}

{\bf Halo Formation by Pancake Instability and Fragmentation} The model
we use to examine the formation of dark-matter dominated halos is that of
cosmological pancake instability and
fragmentation,
previously discussed in detail by \cite{Valinia97}.  
Halos formed by such a pancake instability have density profiles very
similar to those formed hierarchically in CDM models (e.g.
NFW profile \cite{NFW97}), providing a convenient alternative to more
complicated simulations with more realistic initial conditions
\cite{Alvarez00}\cite{Martel01}\cite{ValPhd}.
The ASPH/P$^3$M
simulations considered here were described by \cite{Alvarez00};
this paper extends that analysis to evolutionary trends
in the dark matter halo structure. 

\pagebreak

{\bf Main Results:}
\begin{itemize} 
\item For $a/a_c$ between 3 and 7, the halo can be fit by an
NFW profile, with mass within $r_{200}$ growing linearly with scale factor
$a$, when simulated either with or without gas:
$M_{200}(x)\simeq 0.07x$, where
$x\equiv a/a_c$, and $a_c$ is the scale factor at primary pancake
collapse (see Figs. 1 \& 2). This mass evolution resembles that of
self-similar
spherical infall \cite{Bert85}, despite the anisotropy associated
with pancake collapse and filamentation and periodic boundary
conditions.

\item After $a/a_c=3$, the concentration parameter $c_{NFW}\equiv
r_{200}/r_s$, determined by best-fitting an NFW density profile
\cite{NFW97} to our simulation halos, grows roughly linearly
with scale factor $a$: $c_{NFW}(x)\simeq1.33x-0.18$ (without gas),
$c_{NFW}(x)\simeq1.49x-0.37$ (with gas) (i.e. the linear slope is
steeper in the case with gas included). Fluctuations in $c_{NFW}$ around
this trend are smaller when gas is included. This evolution we find for
$c_{NFW}$ is reminiscent of that reported for halos in CDM
N-body simulations \cite{Bullock00}.  However, the latter applies to halos
of a given mass which are observed at different epochs, and, therefore,
reflects the statistical correlation of halo mass with collapse epoch in
the CDM model, while our result follows an individual halo.

\item The anisotropy parameter $\beta\equiv
1-{\langle}v_t^2{\rangle}/(2{\langle}v_r^2{\rangle})$, where $v_t$($v_r$)
are tangential (radial) velocities, is shown in Figure 2. Pancake halos
are somewhat radially biased, with $\beta \geq 0.6$, about twice the
value reported for halos in CDM N-body simulations
\cite{ENF98}\cite{Thomas98}. With
no gas
included, the average anisotropy in
the halo does not change very much with time, while the inclusion of gas
leads to a slight drop after $a/a_c=5$.
\end{itemize}

\begin{figure}[t!] 
\centerline{\epsfig{file=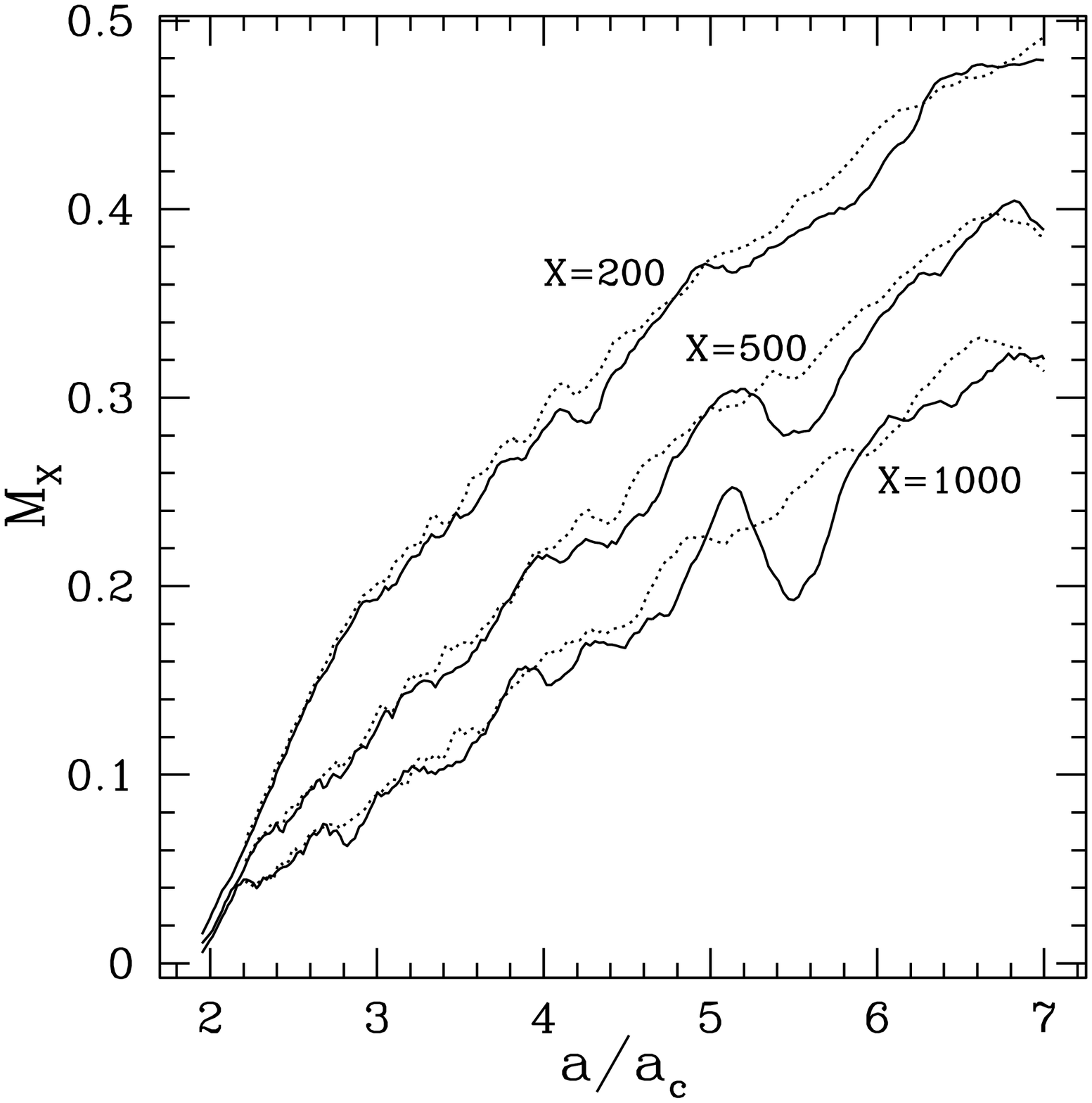,width=2.in}
\epsfig{file=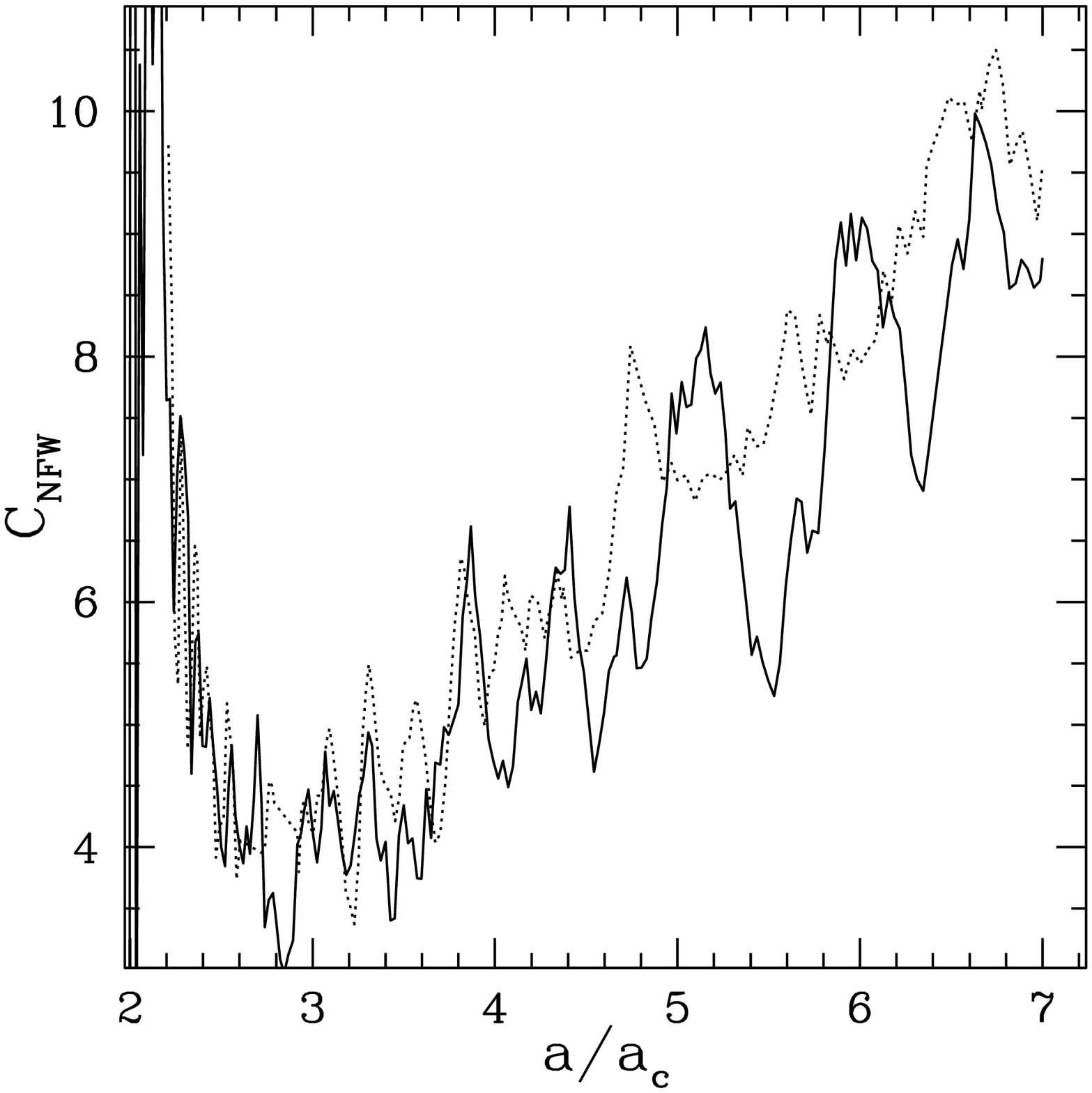,width=2.in}
\epsfig{file=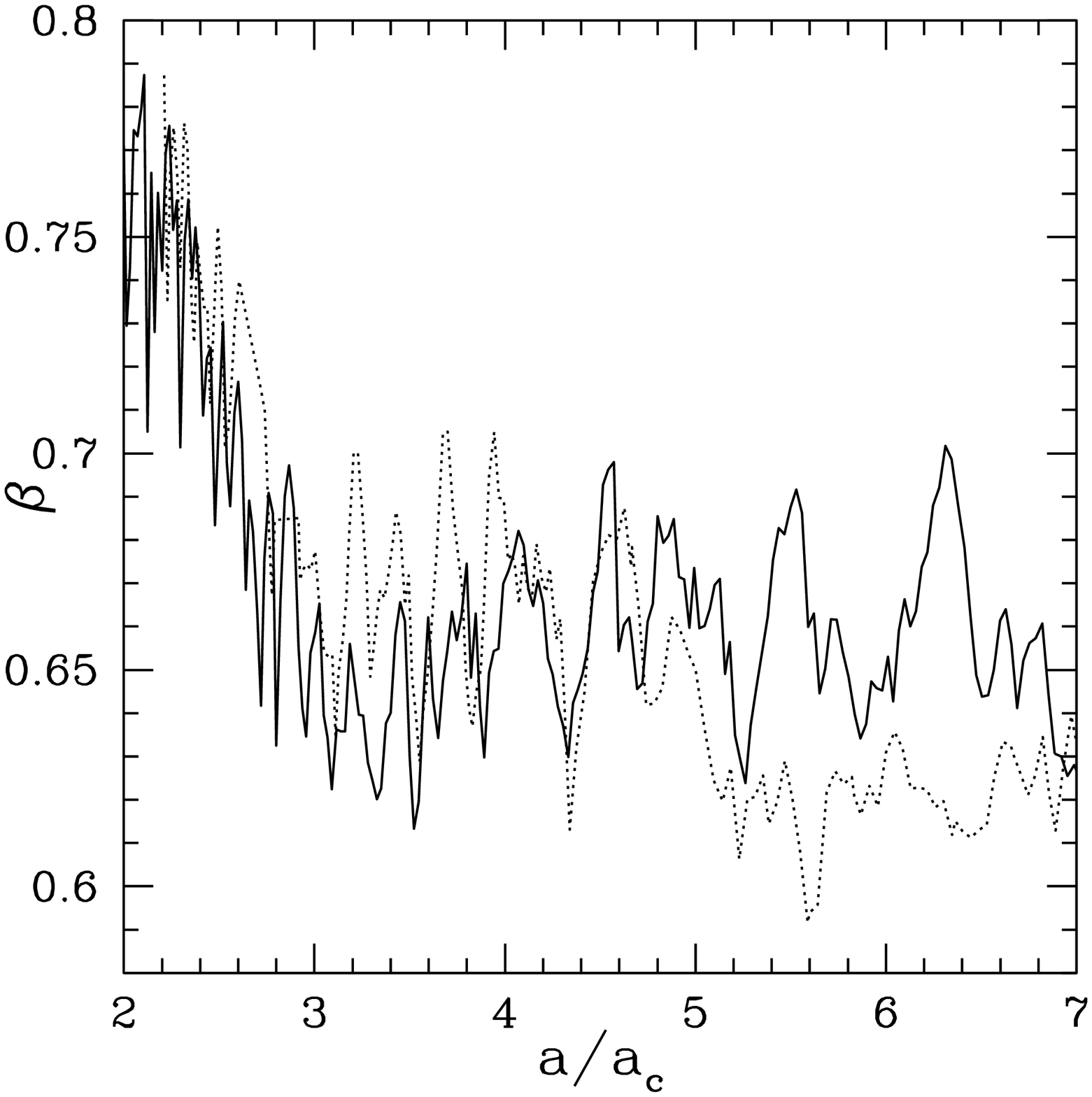,width=2.in}}
\vspace{10pt}

\caption{
(left)
Evolution of halo dark matter integrated mass $M_X$ as simulated with
(dotted) and without (solid) gas, within spheres of average overdensity
$X\equiv{\langle}\rho{\rangle}/\bar{\rho}$ (in computational units, where
$M_{box}=\lambda_p^3\bar{\rho}=1$).  
(middle) Evolution of halo concentration parameter for the dark
matter halo as simulated with (dotted) and without (solid) gas.  
(right) Anisotropy parameter $\beta$
averaged over all dark matter halo particles within a sphere of average
overdensity 200, as simulated with (dotted) and without (solid) gas.}

\label{fig2}
\end{figure}

\pagebreak

This work was supported by NASA ATP grants NAG5-7363 and
NAG5-7821, NSF grant ASC-9504046, and Texas Advanced Research Program
grant 3658-0624-1999.

\end{document}